\documentclass[10pt,a4paper]{article}

\usepackage{hyperref}
\usepackage{settings}
\usepackage{code}

\usepackage{subcaption}

\usepackage[version=0.96]{pgf}
\usepackage{pgfplots}
\usepackage{tikz}

\pgfplotsset{compat=1.5}

\usetikzlibrary{arrows,shapes,snakes,automata,backgrounds,petri,positioning, patterns, arrows.meta, trees}

\pgfplotsset{errorDefaults/.style={
  only marks,
  mark=o,
  mark size=4pt,
  mark options={
    line width=1pt},
  error bars/error bar style={
    line width=1.5pt},
  error bars/error mark options={
    rotate=90,
    mark size=2.5pt,
    line width=1pt},
  error bars/x dir=both,
  error bars/x explicit}}

\pgfplotsset{memoryDefaults/.style={
  only marks,
  mark=o,
  mark size=4pt,
  mark options={
    line width=1pt}}}

\pgfplotsset{
    cycle list={red\\blue\\orange\\black!40!green\\yellow\\green\\pink\\black\\},
}

\newcommand{\timePlotError}[4]{
  \addplot+[errorDefaults] table[x index=#1, y expr=\thisrowno{0}+#2, x error minus index=\number\numexpr#1+1\relax, x error plus index=\number\numexpr#1+2\relax] {#4};
  \addlegendentry{#3}
}

\newcommand{\timePlotErrorNoNodes}[4]{
  \addplot+[errorDefaults, nodes near coords={}] table[x index=#1, y expr=\thisrowno{0}+#2, x error minus index=\number\numexpr#1+1\relax, x error plus index=\number\numexpr#1+2\relax] {#4};
  \addlegendentry{#3}
}

\newcommand{\memPlot}[4]{
  \addplot+[memoryDefaults] table[x index=#1, y expr=\thisrowno{0}+#2] {#4};
  \addlegendentry{#3}
}

\newcommand{\memPlotNoNodes}[4]{
  \addplot+[memoryDefaults, nodes near coords={}] table[x index=#1, y expr=\thisrowno{0}+#2] {#4};
  \addlegendentry{#3}
}

\title{Integrating Enzyme-generated functions into CoDiPack}
\author{M. Sagebaum\footnote{Corresponding author: max.sagebaum@scicomp.uni-kl.de}, M. Aehle, N.R. Gauger \\ \ \\
\small Chair for Scientific Computing\\ \small University of Kaiserslautern-Landau (RPTU), Germany}

\date {\today}

\begin{document}

\maketitle
\begin{abstract}
In operator overloading algorithmic differentiation, it can be beneficial to create custom derivative functions for some parts of the code base. For manual implementations of the derivative functions, it can be quite cumbersome to derive, implement, test, and maintain these. The process can be automated with source transformation algorithmic differentiation tools like Tapenade or compiler-based algorithmic differentiation tools like Enzyme. This eliminates most of the work required from a manual implementation but usually has the same efficiency with respect to timing and memory. We present a new helper in CoDiPack that allows Enzyme-generated derivative functions to be automatically added during the recording process of CoDiPack. The validity of the approach is demonstrated on a synthetic benchmark, which shows promising results.
\end{abstract}

\section{Introduction}
The creation of the LLVM infrastructure \cite{LLVM:CGO04} facilitates the development of source analysis and transformation tools. The Enzyme project \cite{NEURIPS2020_9332c513} uses this infrastructure to apply algorithmic differentiation (AD) \cite{grie08} to source code at the compiler level. AD describes how computer programs can be modified to automatically compute derivatives alongside the original (also known as primal) computation. For a function $y = F(x)$ with $x \in \R^m$ and $y \in \R^n$, the forward mode of AD computes
\begin{equation}
  \dot y = \frac{d F}{d x}(x) \dot x
  \label{eq:adForward}
\end{equation}
where $\dot x \in \R^m$ is the tangent seeding and $\dot y \in \R^n$ is the derivative result. The reverse mode of AD computes
\begin{equation}
  \bar x = \frac{d F}{d x}^\top(x) \bar y
  \label{eq:adReverse}
\end{equation}
where $\bar y \in \R^n$ is the adjoint seeding and $\bar x \in \R^m$ is the derivative result. Both modes do not set up the Jacobian matrix. Rather, they compute the derivatives by applying the chain rule and directional derivative on a statement-by-statement level.

Traditionally, AD is applied to a computer program either via operator overloading or source transformation. The operator overloading approach exchanges the computational type like \ic{double} with a so-called active type of the AD tool implementation like \ic{adouble} for ADOL-C \cite{Walther2012Gsw} or \ic{codi::RealReverse} for CoDiPack \cite{SaAlGauTOMS2019}.
In the reverse mode of AD, operator overloading tools record a tape storing all the necessary information for evaluating Equation \eqref{eq:adReverse}. The data on the tape can be thought of as the computational graph of the program. The tape is then interpreted in a reverse manner to compute the reverse mode of AD. Source transformation tools, like Tapenade \cite{Hascoet2013TTA} or OpenAD \cite{Naumann2006ACb}, apply AD by parsing the source code of the program and generating a new code extending it by the additional AD computations.

Operator overloading AD is usually applied to legacy software projects where the original software design did not include AD. Source transformation AD on legacy software usually has problems analyzing the computational dependencies on a global level, which makes it quite cumbersome to apply. On the other hand, source transformation is usually used for new software projects where the software design includes AD. Here, the derivative code generated by source transformation is usually more efficient in terms of memory and computational time than the derivative computation done by operator overloading AD tools.

The approach of Enzyme is similar to source transformation. Enzyme works on the intermediate representation (IR) of LLVM, a language-independent high-level assembly language. It contains enough information such that AD can still be applied without any drawbacks. By using the IR, Enzyme becomes a multi-language AD tool and does not have the problem that it needs to parse and understand the code constructs of the original language. Because Enzyme is similar to source transformation, it inherits the usual drawbacks. Applying Enzyme to a large code base can be cumbersome and a full dependency analysis might not be possible.

A combination of both approaches can be beneficial for the overall runtime and memory consumption. First, a code base can be differentiated by using an operator overloading AD tool. Afterward, the computational hot spots can be handled by Enzyme and integrated into the tape of the operator overloading tool. The derivative functions for the computational hot spots could also be implemented by hand. In doing so, an extra development effort for the implementation, testing, and debugging is required which needs to be repeated every time the original function is changed. Applying Enzyme or a source transformation tool on those hot spots is therefore much more efficient regarding development time.

In this report, we present an extension to CoDiPack which provides a very simple way to add the generated derivative functions from Enzyme to the CoDiPack tapes. The next section will introduce the approach and afterward in Section \ref{sec:dataManagement} the memory footprint of external functions is explained in more detail. Section \ref{sec:testCase} describes the Burgers' test case for the performance measurements. The results are discussed in Section \ref{sec:results}.

\section{Adding Enzyme to CoDiPack}

An integration of Enzyme into a CoDiPack tape is done by so-called external functions. They are called by the tape during the reverse interpretation and are usually used to handle libraries for which the AD tool can not be applied. In the case of Enzyme, they are used to improve the memory and evaluation time of certain code parts.

Adding an external function to CoDiPack can either be done by using the tape interface or by using the \ic{ExternalFunctionHelper} structure. The latter assumes that the function has the layout
\begin{codeRefInline}{code:funcDef}{Function definition for external functions.}
void func(double const* x, size_t m, double* y, size_t n,
          ExternalFunctionUserData* d)
\end{codeRefInline}
where $x \in \R^m$ is the input, $y \in \R^n$ is the output, and \ic{d} is additional data for the function.
Based on the definition in Listing \ref{code:funcDef}, the user has to implement the forward and reverse mode implementation of the function. Following the regular naming conventions of AD, the layout for the functions are:
\begin{codeRefInline}{code:funcDerivDef}{Forward and reverse mode function definition for external functions.}
void func_d(double const* x, double const* x_d, size_t m,
            double* y, double* y_d, size_t n,
            ExternalFunctionUserData* d);
void func_b(double const* x, double* x_b, size_t m,
            double const* y, double const* y_b, size_t n,
            ExternalFunctionUserData* d);
\end{codeRefInline}

Instead of implementing \ic{func_d} and \ic{func_b} manually, with the mentioned disadvantage, Enzyme can be used to automate the generation of the two functions. The generator conventions for Enzyme \cite{NEURIPS2020_9332c513} will actually generate the definitions in Listing \ref{code:funcDerivDef} when applied to Listing \ref{code:funcDef}. See Appendix \ref{app:enzymeDefinitions} for the Enzyme generation code.

This observation allows for a specialized implementation of the CoDiPack \ic{ExternalFunctionHelper} for Enzyme. The \ic{EnzymeExternalFunctionHelper} adds an external function just by providing the function as a template argument. Listing \ref{code:enzymeSmallExample} shows an example usage of the new helper. Here, the function \ic{vecPow2} is added to the tape. It raises each element in the vector to the power of two and writes it into the output vector. See Appendix \ref{app:enzymeFullExample} for a full example.

\begin{codeRef}{code:enzymeSmallExample}{Example of adding an external function with Enzyme-generated derivatives.}
  codi::EnzymeExternalFunctionHelper<codi::RealReverse> eh;
  eh.template addToTape<vecPow2>(x, size, y, size);
\end{codeRef}

\section{Data management of external functions}
\label{sec:dataManagement}

In this section we want to take a closer look at the data management of external functions in CoDiPack. For optimal performance, it is critical to look at the actual memory size an external function requires on the tape. Functions that are too small will require more memory than the taped version and therefore decrease performance.

\newcommand{\MEM}{\text{MEM}}
The memory required by an external function in CoDiPack can be calculated as
\begin{equation}
  \begin{aligned}
    \MEM(ext\_func(func)) = &\ \MEM(\text{overhead}) \\
                           & + m \cdot (\MEM(double) + \MEM(int))\\
                           & + n \cdot \MEM(int)\\
  \end{aligned}
\end{equation}
where $m \in \N$ is the number of inputs and $n \in \N$ the number of outputs. CoDiPack stores the primal value and the identifier of input AD values. For outputs only the identifier is stored. $\MEM(\text{overhead})$ varies for the different tape implementations in CoDiPack but is around 256 bytes. It consists of position information (varies) and the data structure for the external function helper (fixed).
There are some special cases where the above equation can be modified:
\begin{itemize}
  \item \textbf{Result $y$:} It can be beneficial to store the primal values of the outputs. This will add another $n \cdot \MEM(double)$ bytes to the memory.
  \item \textbf{Primal value tapes:} Here, the primal values are available in the primal value vector of the tape and do not need to be stored. This removes $m \cdot \MEM(double)$ from the above equation.
\end{itemize}

The overhead of around $256$ bytes prohibits the use of external functions for smaller functions. An additional large overhead can occur when the same inputs are stored for multiple external functions. It is therefore advisable to use external functions where a large amount of data is handled.

\section{Burgers' test case}
\label{sec:testCase}

The coupled Burgers' equation is an established test case for the performance comparison of CoDiPack implementations and is described in \cite{SaAlGauTOMS2019} and \cite{SaAlGa2018OMS}. We want to use the same test case for the performance evaluations in this report.
For completeness, we recapitulate the problem formulation here.

The coupled Burgers' equation \cite{biazar2009exact,bahadir2003fully,zhu2010numerical}
\begin{align}
  u_t + uu_x + vu_y &= \frac{1}{R}(u_{xx} + u_{yy}), \\
  v_t + uv_x + vv_y &= \frac{1}{R}(v_{xx} + v_{yy})
\end{align}
is discretized with an upwind finite difference scheme.
The initial and boundary conditions are taken from the exact solution
\begin{align}
  u(x, y, t) &= \frac{x + y - 2xt}{1 - 2t^2} \quad (x,y,t) \in D \times \R,\\
  v(x, y, t) &= \frac{x - y - 2yt}{1 - 2t^2} \quad (x,y,t) \in D \times \R
\end{align}
given in \cite{biazar2009exact}.
The computational domain $D$ is the unit square $D = [0,1] \times [0,1] \subset \R \times \R$.
As far as the differentiation is concerned, we choose the initial solution of the time stepping scheme as input parameters, and as output parameter we take the norm of the final solution.

The test case is extended for this report such that Enzyme is applied on one time update step. The differentiated function consists of the 2D loop over the domain for the state update of $u$ and $v$. A code example is given in Appendix \ref{app:enzymeUpdateLoop}. For a classification of the Enzyme results, Tapenade is also applied on the same loop. The example is chosen such that no data is written to the stack by Tapenade or Enzyme, which does not add any additional load on the memory bandwidth.

The node for the test case consists of two Intel Xeon 6126 CPUs with a total of 24 cores and 384 GB of main memory.
The computational grid contains $601\times 601$ elements and is solved for 16 time iterations.
The code is compiled with clang version 14. We remark that similar results are obtained on nodes with Epyc and Haswell CPUs.
All time measurements are averaged over 20 evaluations.

For the time measurements two different configurations are tested:
\begin{itemize}
  \item The \emph{multi} test configuration runs the same process on each of the 24 cores, which simulates a use case where the full memory bandwidth of the socket is used.
  \item The \emph{single} test configuration runs just one process on the whole node, which eliminates the memory bandwidth limitations and provides a better view on the computational performance.
\end{itemize}

\section{Results}
\label{sec:results}

We compare the runtime and memory for the four major CoDiPack tape implementations:
\begin{itemize}
  \item \textbf{Jacobian linear} - Jacobian taping approach \cite{SaAlGauTOMS2019} with a linear index management \cite{sagebaum2021index}
  \item \textbf{Jacobian reuse} - Jacobian taping approach with a reuse index management \cite{sagebaum2021index} including copy optimizations
  \item \textbf{primal linear} - Primal value taping approach \cite{SaAlGa2018OMS} with a linear index management
  \item \textbf{primal reuse} - Primal value taping approach with a reuse index management including copy optimizations
\end{itemize}
For the primal value taping approaches, we enable the specialized handling mentioned in Section \ref{sec:dataManagement} (\emph{handling=on}). Here, the primal values are recovered from the primal value vector of the tape and not stored in the external function data.

\begin{figure}
  \center
  \begin{tikzpicture}
    \begin{axis}[
      height=6cm,
      width=0.9\textwidth,
      xlabel={Memory in MB},
      xmin=0,
      ymax=0.1,
      ytick={-1, -2, -3, -4, -5, -6},
      yticklabels={Jacobian linear, Jacobian index, Primal reuse handling=off, Primal reuse handling=off, Primal linear handling=on, Primal reuse handling=on},
      legend style={at={(1.,1.)},anchor=south east},
      nodes near coords,
      nodes near coords style={yshift=3pt, color=black},
      point meta=rawx
    ]
      \memPlot{13}{0.100}{No external functions}{results/codi2_withImprovements_1.dat}
      \memPlot{13}{0.000}{Enzyme-generated derivative}{results/codi2_enzymeGridLoop_withImprovements_1.dat}
      \memPlotNoNodes{13}{-0.100}{Tapenade-generated derivative}{results/codi2_tapenadeGridLoop_withImprovements_1.dat}
    \end{axis}
  \end{tikzpicture}
  \caption{Recorded tape memory. (Does not include the data of the external functions.)}
  \label{fig:memoryTape}
\end{figure}
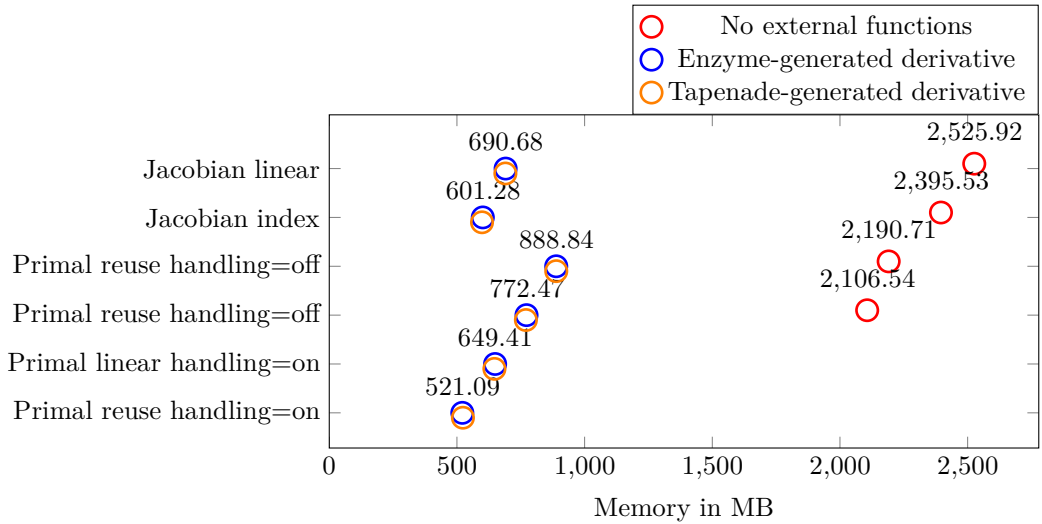

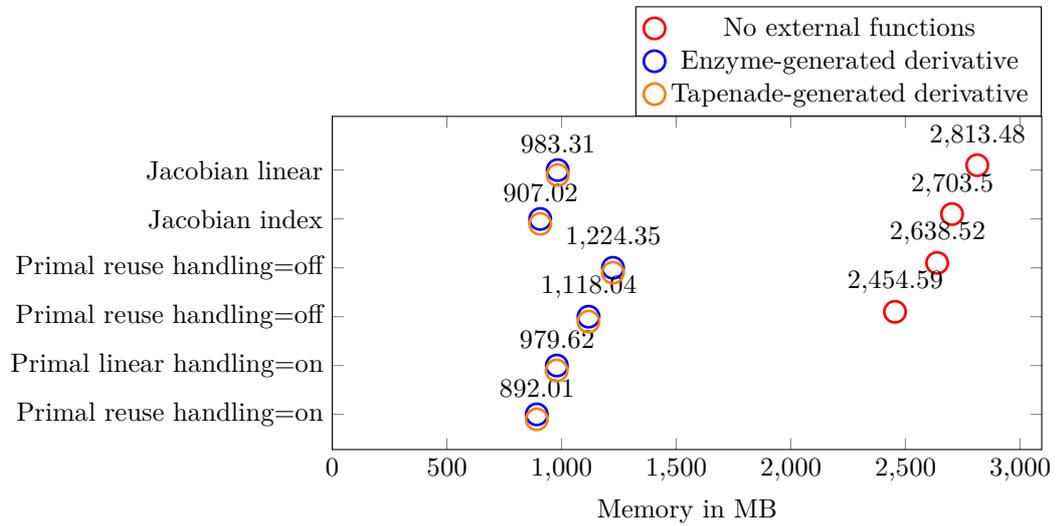
\begin{figure}
  \center
  \begin{tikzpicture}
    \begin{axis}[
      height=6cm,
      width=0.9\textwidth,
      xlabel={Memory in MB},
      xmin=0,
      ymax=0.1,
      ytick={-1, -2, -3, -4, -5, -6},
      yticklabels={Jacobian linear, Jacobian index, Primal reuse handling=off, Primal reuse handling=off, Primal linear handling=on, Primal reuse handling=on},
      legend style={at={(1.,1.)},anchor=south east},
      nodes near coords,
      nodes near coords style={yshift=3pt, color=black},
      point meta=rawx
    ]
      \memPlot{14}{0.100}{No external functions}{results/codi2_withImprovements_1.dat}
      \memPlot{14}{0.000}{Enzyme-generated derivative}{results/codi2_enzymeGridLoop_withImprovements_1.dat}
      \memPlotNoNodes{14}{-0.100}{Tapenade-generated derivative}{results/codi2_tapenadeGridLoop_withImprovements_1.dat}
    \end{axis}
  \end{tikzpicture}
  \caption{Memory high watermark.}
  \label{fig:memoryHighWatermark}
\end{figure}

The memory results in Figure \ref{fig:memoryTape} and \ref{fig:memoryHighWatermark} show a significant reduction in the tape and overall memory when the loop is handled either by Enzyme or Tapenade. The tape memory in Figure \ref{fig:memoryTape} does not include the data stored by the external functions, therefore the high water mark is better suited to judge the general savings in memory. The Jacobian tapes have a lower memory footprint after the loop handling, because it is more expensive for the primal value tapes to manage external function outputs. Memory for the primal values tapes can be improved by recovering the primal values from the tape. With this option the memory of the Jacobian tapes and primal value tapes are nearly the same.

\begin{figure}
  \center
  \begin{tikzpicture}
    \begin{axis}[
      height=6cm,
      width=0.9\textwidth,
      xlabel={Time in seconds},
      xmin=0,
      ymax=0.1,
      ytick={-1, -2, -3, -4, -5, -6},
      yticklabels={Jacobian linear, Jacobian index, Primal reuse handling=off, Primal reuse handling=off, Primal linear handling=on, Primal reuse handling=on},
      legend style={at={(1.,1.)},anchor=south east},
      nodes near coords,
      nodes near coords style={yshift=3pt, color=black},
      point meta=rawx
    ]
      \timePlotError{4}{0.100}{No external functions}{results/codi2_withImprovements_1.dat}
      \timePlotError{4}{0.000}{Enzyme-generated derivative}{results/codi2_enzymeGridLoop_withImprovements_1.dat}
      \timePlotErrorNoNodes{4}{-0.100}{Tapenade-generated derivative}{results/codi2_tapenadeGridLoop_withImprovements_1.dat}
    \end{axis}
  \end{tikzpicture}
  \caption{Reversal time for the single configuration.}
  \label{fig:timingReverseSingle}
\end{figure}
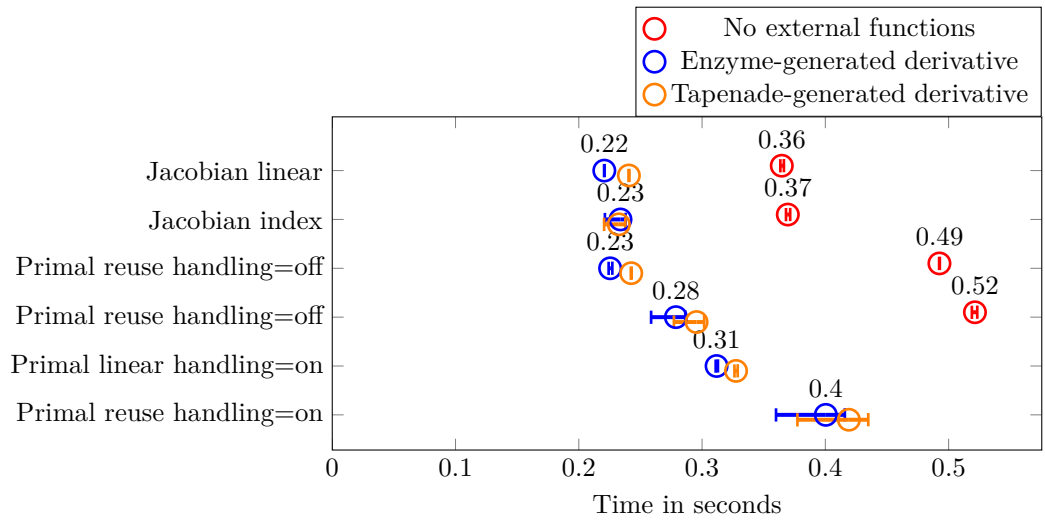

The timing results in Figure \ref{fig:timingRecordSingle} show the recording time for the different configurations. Handling the 2D loop with Enzyme or Tapenade reduces the time by around 45 \% which is mostly due to the reduction in tape memory. The memory bandwidth for writing the data to the tape is usually the limiting factor. The reversal times in Figure \ref{fig:timingReverseSingle} show the same picture and an improvement by 35 \% can be seen. The degradation in case of the primal value tapes for \emph{handling=on} is due to the recovery of the primal values from the tape. This is a random access on the primal value vector which can be quite slow.

\begin{figure}
  \center
  \begin{tikzpicture}
    \begin{axis}[
      height=6cm,
      width=0.9\textwidth,
      xlabel={Time in seconds},
      xmin=0,
      ymax=0.1,
      ytick={-1, -2, -3, -4, -5, -6},
      yticklabels={Jacobian linear, Jacobian index, Primal reuse handling=off, Primal reuse handling=off, Primal linear handling=on, Primal reuse handling=on},
      legend style={at={(1.,1.)},anchor=south east},
      nodes near coords,
      nodes near coords style={yshift=3pt, color=black},
      point meta=rawx
    ]
      \timePlotError{1}{0.100}{No external functions}{results/codi2_withImprovements_1.dat}
      \timePlotError{1}{0.000}{Enzyme-generated derivative}{results/codi2_enzymeGridLoop_withImprovements_1.dat}
      \timePlotErrorNoNodes{1}{-0.100}{Tapenade-generated derivative}{results/codi2_tapenadeGridLoop_withImprovements_1.dat}
    \end{axis}
  \end{tikzpicture}
  \caption{Recording time for the single configuration.}
  \label{fig:timingRecordSingle}
\end{figure}
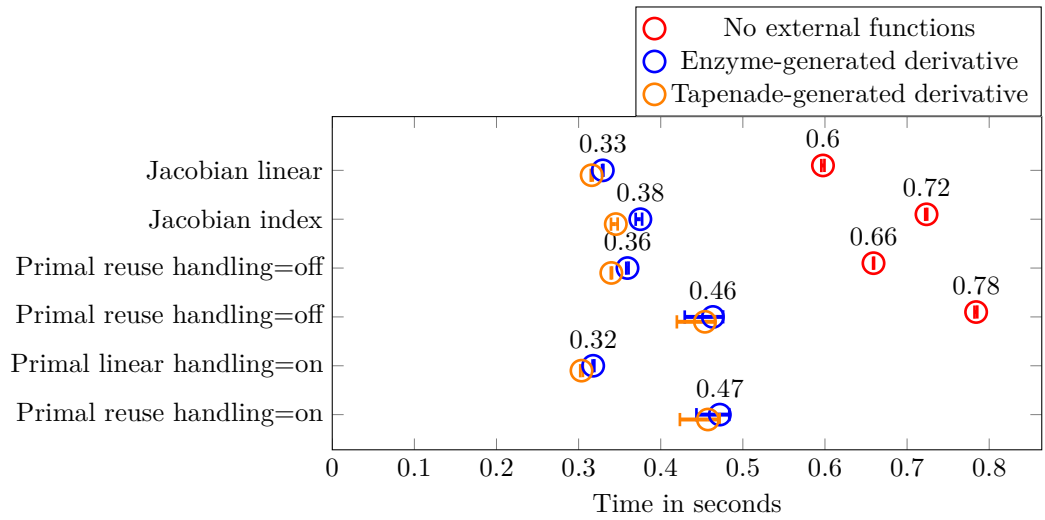

\begin{figure}
  \center
  \begin{tikzpicture}
    \begin{axis}[
      height=6cm,
      width=0.9\textwidth,
      xlabel={Time in seconds},
      xmin=0,
      ymax=0.1,
      ytick={-1, -2, -3, -4, -5, -6},
      yticklabels={Jacobian linear, Jacobian index, Primal reuse handling=off, Primal reuse handling=off, Primal linear handling=on, Primal reuse handling=on},
      legend style={at={(1.,1.)},anchor=south east},
      nodes near coords,
      nodes near coords style={yshift=3pt, color=black},
      point meta=rawx
    ]
      \timePlotError{4}{0.100}{No external functions}{results/codi2_withImprovements_24.dat}
      \timePlotError{4}{0.000}{Enzyme-generated derivative}{results/codi2_enzymeGridLoop_withImprovements_24.dat}
      \timePlotErrorNoNodes{4}{-0.100}{Tapenade-generated derivative}{results/codi2_tapenadeGridLoop_withImprovements_24.dat}
    \end{axis}
  \end{tikzpicture}
  \caption{Reversal time for the multi configuration.}
  \label{fig:timingReverseMulti}
\end{figure}
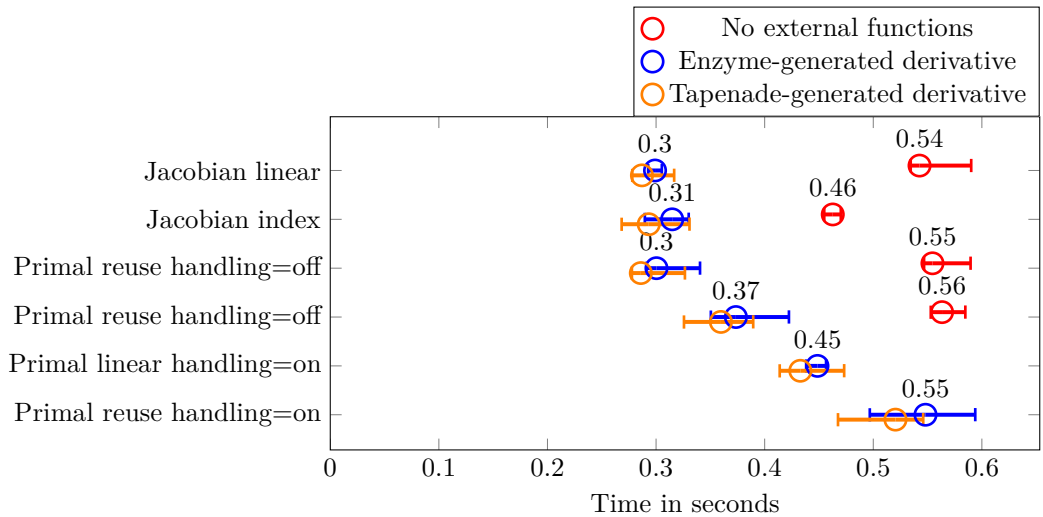

The timing results for the multi-configuration in Figure \ref{fig:timingRecordMulti} and Figure \ref{fig:timingReverseMulti} show the same tendencies as the single results. For the recording, the timing reduction is around 32 \%. All 24 cores are using the memory bandwidth for creating the external function which makes its limitation effect more pronounced in this configuration. The reversal has now a reduction of about 45 \% in the timing which is actually better than in the single configuration. The memory bandwidth limitations are more pronounced for the large tapes with no external functions. Also, the structured data access of the generated derivative functions helps to recover more of the peak performance from the CPU.

\begin{figure}
  \center
  \begin{tikzpicture}
    \begin{axis}[
      height=6cm,
      width=0.9\textwidth,
      xlabel={Time in seconds},
      xmin=0,
      ymax=0.1,
      ytick={-1, -2, -3, -4, -5, -6},
      yticklabels={Jacobian linear, Jacobian index, Primal reuse handling=off, Primal reuse handling=off, Primal linear handling=on, Primal reuse handling=on},
      legend style={at={(1.,1.)},anchor=south east},
      nodes near coords,
      nodes near coords style={yshift=3pt, color=black},
      point meta=rawx
    ]
      \timePlotError{1}{0.100}{No external functions}{results/codi2_withImprovements_24.dat}
      \timePlotError{1}{0.000}{Enzyme-generated derivative}{results/codi2_enzymeGridLoop_withImprovements_24.dat}
      \timePlotErrorNoNodes{1}{-0.100}{Tapenade-generated derivative}{results/codi2_tapenadeGridLoop_withImprovements_24.dat}
    \end{axis}
  \end{tikzpicture}
  \caption{Recording time for the multi configuration.}
  \label{fig:timingRecordMulti}
\end{figure}
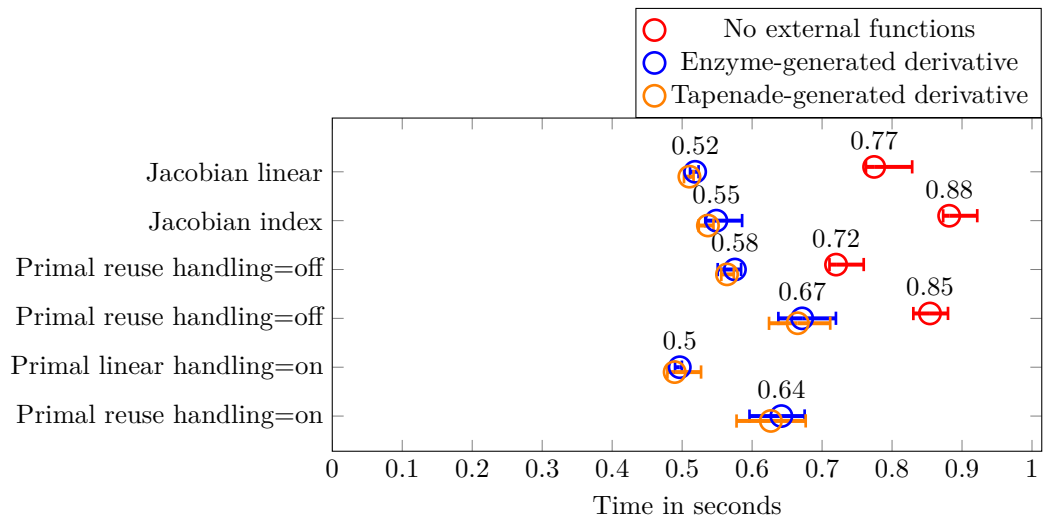

\section{Conclusion}
In this study, we demonstrated how Enzyme-generated derivative functions can be included in CoDiPack. The validity of the approach was demonstrated on a numerical benchmark where the memory was reduced from 2.8 GB to 1.0 GB. The recording time was improved by 32 \% and the reversal time by 45 \% in the memory bandwidth-limited case. These results confirm the validity of the approach and further studies need to be conducted on real-world problems.

\section{Acknowledgments}

We would like to thank William Moses for helping us to apply Enzyme to our test problem.

\appendix

\section{Enzyme function generation}
\label{app:enzymeDefinitions}
Forward and reverse AD mode generation for an external function with Enzyme.
\begin{code}
template<PrimalFunc func>
void enzymeDiff_d(double const* x, double const* x_d, size_t m,
                 double * y, double * y_d, size_t n,
                 ExternalFunctionUserData* d) {
  __enzyme_fwddiff(
      (void*) func,
      enzyme_dup, x, x_d,
      enzyme_const, m,
      enzyme_dup, y, y_d,
      enzyme_const, n,
      enzyme_const, d); 
}

template<PrimalFunc func>
void enzymeDiff_b(double const* x, double* x_b, size_t m,
                 double const* y, double const* y_b, size_t n,
                 ExternalFunctionUserData* d) {
  __enzyme_autodiff(
      (void*) func,
      enzyme_dup, x, x_b,
      enzyme_const, m,
      enzyme_dup, y, y_b,
      enzyme_const, n,
      enzyme_const, d); 
}   
\end{code}

\section{Full \ic{EnzymeExternalFunctionHelper} example}
\label{app:enzymeFullExample}
\begin{code}
#include <codi.hpp>
#include <iostream>

using Real = codi::RealReverse;
using Tape = typename Real::Tape;
using BaseReal = typename Real::Real;

void vecPow2(const BaseReal* x, size_t m, BaseReal* y,
             size_t n, codi::ExternalFunctionUserData* d) {
  for(size_t i = 0; i < m; i += 1) {
    y[i] = x[i] * x[i];
  }
}

int main(int nargs, char** args) {
  Tape& tape = Real::getTape();
  size_t size = 10;

  Real* x = new Real[size];
  Real* y = new Real[size];
  for(size_t i = 0; i < size; i += 1) {
    x[i] = BaseReal(i + 1);
  }
  
  tape.setActive();
  for(size_t i = 0; i < size; i += 1) { tape.registerInput(x[i]); }

  codi::EnzymeExternalFunctionHelper<codi::RealReverse> eh;
  eh.template addToTape<vecPow2>(x, size, y, size);

  for(size_t i = 0; i < size; i += 1) { tape.registerOutput(y[i]); }

  tape.setPassive();
  for(size_t i = 0; i < size; i += 1) { y[i].setGradient(1.0); }

  tape.evaluate();

  std::cout << "Solution y:" << std::endl;
  for(size_t i = 0; i < size; i += 1) {
    std::cout << "  " << i << " : " << y[i] << std::endl;
  }
  std::cout << "Adjoint x:" << std::endl;
  for(size_t i = 0; i < size; i += 1) {
    std::cout << "  " << i << " : " << x[i].gradient() << std::endl;
  }

  tape.reset();

  return 0;
}
\end{code}

\section{Enzyme handling of \ic{updateField}}
\label{app:enzymeUpdateLoop}

\begin{code}
void updateFieldExtFunc(double const* x, size_t m, double* y, size_t n,
                        codi::ExternalFunctionUserData* d) {

  size_t xSize = d->getData<size_t>();
  size_t ySize = d->getData<size_t>();
  double dTbyDX = d->getData<double>();
  double dTbyDX2 = d->getData<double>();
  double oneOverR = d->getData<double>();
  size_t inputGridSize = (xSize + 2) * (ySize + 2);
  size_t outputGridSize = xSize * ySize;

  double const* u_t = &x[0];
  double const* v_t = x + inputGridSize;
  double* u_tp = y;
  double* v_tp = y + outputGridSize;

  for (int yPos = 0; yPos < ySize; yPos += 1) {
    for (int xPos = 0; xPos < xSize; xPos += 1) {
      size_t index_out = getArrayPos(xPos, yPos, xSize);
      size_t index = getArrayPos(xPos + 1, yPos + 1, xSize + 2);
      size_t index_xp = index + 1;
      size_t index_xm = index - 1;
      size_t index_yp = index + xSize + 2;
      size_t index_ym = index - xSize - 2;

      updateElement(u_tp, u_t, u_t, v_t, index_out, index, index_xp,
                    index_xm, index_yp, index_ym, dTbyDX, dTbyDX2,
                    oneOverR);
      updateElement(v_tp, v_t, u_t, v_t, index_out, index, index_xp,
                    index_xm, index_yp, index_ym, dTbyDX, dTbyDX2,
                    oneOverR);
    }
  }
}
  
using Number = codi::RealReverse;

inline void updateFieldWithEnzyme(int xStart, int xEnd,
                                  int yStart, int yEnd,
                                  Number *u_tp, Number const* u_t,
                                  Number *v_tp, Number const *v_t) {
  codi::EnzymeExternalFunctionHelper<Number> helper;
  
  size_t xSize = xEnd - xStart + 1;
  size_t ySize = yEnd - yStart + 1;

  for (int yPos = yStart - 1; yPos <= yEnd + 1; yPos += 1) {
    for (int xPos = xStart - 1; xPos <= xEnd + 1; xPos += 1) {
      size_t index = getArrayPos(xPos, yPos, xSize);
      helper.addInput(u_t[index]);
    }
  }
  for (int yPos = yStart - 1; yPos <= yEnd + 1; yPos += 1) {
    for (int xPos = xStart - 1; xPos <= xEnd + 1; xPos += 1) {
      size_t index = getArrayPos(xPos, yPos, xSize);
      helper.addInput(v_t[index]);
    }
  }
  for (int yPos = yStart; yPos <= yEnd; yPos += 1) {
    for (int xPos = xStart; xPos <= xEnd; xPos += 1) {
      size_t index = getArrayPos(xPos, yPos, xSize);
      helper.addOutput(u_tp[index]);
    }
  }
  for (int yPos = yStart; yPos <= yEnd; yPos += 1) {
    for (int xPos = xStart; xPos <= xEnd; xPos += 1) {
      size_t index = getArrayPos(xPos, yPos, xSize);
      helper.addOutput(v_tp[index]);
    }
  }

  codi::ExternalFunctionUserData& userData = 
      helper.getExternalFunctionUserData();
  userData.addData(xSize);
  userData.addData(ySize);
  userData.addData(props.dTbyDX);
  userData.addData(props.dTbyDX2);
  userData.addData(props.oneOverR);

  helper.template addToTape<updateFieldExtFunc>();
}
\end{code}

\bibliographystyle{alphaurl}
\bibliography{literature}

\newcommand{\etalchar}[1]{$^{#1}$}
\begin{thebibliography}{NUW{\etalchar{+}}06}

\bibitem[BA09]{biazar2009exact}
J.~Biazar and H.~Aminikhah.
\newblock {Exact and numerical solutions for non-linear Burgers' equation by
  VIM}.
\newblock {\em Mathematical and Computer Modelling}, 49(7):1394--1400, 2009.
\newblock \href {https://doi.org/10.1016/j.mcm.2008.12.006}
  {\path{doi:10.1016/j.mcm.2008.12.006}}.

\bibitem[Bah03]{bahadir2003fully}
A.~Bahad{\i}r.
\newblock {A fully implicit finite-difference scheme for two-dimensional
  Burgers' equations}.
\newblock {\em Applied Mathematics and Computation}, 137(1):131--137, 2003.
\newblock \href {https://doi.org/10.1016/S0096-3003(02)00091-7}
  {\path{doi:10.1016/S0096-3003(02)00091-7}}.

\bibitem[GW08]{grie08}
A.~Griewank and A.~Walther.
\newblock {\em {Evaluating Derivatives}}.
\newblock Society for Industrial and Applied Mathematics, second edition, 2008.
\newblock \href {https://doi.org/10.1137/1.9780898717761}
  {\path{doi:10.1137/1.9780898717761}}.

\bibitem[HP13]{Hascoet2013TTA}
L.~Hasco{\"e}t and V.~Pascual.
\newblock The {T}apenade automatic differentiation tool: {P}rinciples, model,
  and specification.
\newblock {\em {ACM} Transactions on Mathematical Software}, 39(3):20:1--20:43,
  2013.
\newblock \href {https://doi.org/10.1145/2450153.2450158}
  {\path{doi:10.1145/2450153.2450158}}.

\bibitem[LA04]{LLVM:CGO04}
C.~Lattner and V.~Adve.
\newblock {LLVM: A Compilation Framework for Lifelong Program Analysis \&
  Transformation}.
\newblock In {\em {Proceedings of the 2004 International Symposium on Code
  Generation and Optimization (CGO'04)}}, Palo Alto, California, Mar 2004.
\newblock URL: \url{https://llvm.org/pubs/2004-01-30-CGO-LLVM.pdf}.

\bibitem[NUW{\etalchar{+}}06]{Naumann2006ACb}
U.~Naumann, J.~Utke, C.~Wunsch, C.~Hill, P.~Heimbach, M.~Fagan, N.~Tallent, and
  M.~Strout.
\newblock {Adjoint Code by Source Transformation with {O}pen{AD/F}}.
\newblock In Pieter Wesseling, Jacques P\'eriaux, and Eugenio O{\~n}ate,
  editors, {\em Proceedings of the European Conference on Computational Fluid
  Dynamics (ECCOMAS CFD 2006)}. TU Delft, 2006.
\newblock URL:
  \url{http://proceedings.fyper.com/eccomascfd2006/documents/35.pdf}.

\bibitem[SAG18]{SaAlGa2018OMS}
M.~Sagebaum, T.~Albring, and N.R. Gauger.
\newblock {Expression templates for primal value taping in the reverse mode of
  algorithmic differentiation}.
\newblock {\em Optimization Methods and Software}, 33(4-6):1207--1231, 2018.
\newblock \href {https://doi.org/10.1080/10556788.2018.1471140}
  {\path{doi:10.1080/10556788.2018.1471140}}.

\bibitem[SAG19]{SaAlGauTOMS2019}
M.~Sagebaum, T.~Albring, and N.R. Gauger.
\newblock {High-Performance Derivative Computations Using CoDiPack}.
\newblock {\em ACM Trans. Math. Softw.}, 45(4), December 2019.
\newblock \href {https://doi.org/10.1145/3356900} {\path{doi:10.1145/3356900}}.

\bibitem[SBG21]{sagebaum2021index}
M.~Sagebaum, J.~Blühdorn, and N.R. Gauger.
\newblock {Index handling and assign optimization for Algorithmic
  Differentiation reuse index managers}.
\newblock arXiv cs.MS 2006.12992, 2021.
\newblock URL: \url{https://arxiv.org/abs/2006.12992}.

\bibitem[WC20]{NEURIPS2020_9332c513}
W.~S. William and V.~Churavy.
\newblock Instead of rewriting foreign code for machine learning, automatically
  synthesize fast gradients.
\newblock In {\em Proceedings of the 34th International Conference on Neural
  Information Processing Systems}, NIPS'20, Red Hook, NY, USA, 2020. Curran
  Associates Inc.
\newblock \href {https://doi.org/10.5555/3495724.3496770}
  {\path{doi:10.5555/3495724.3496770}}.

\bibitem[WG12]{Walther2012Gsw}
A.~Walther and A.~Griewank.
\newblock {Getting started with ADOL-C}.
\newblock In U.~Naumann and O.~Schenk, editors, {\em Combinatorial Scientific
  Computing}, chapter~7, pages 181--202. Chapman-Hall CRC Computational
  Science, 2012.
\newblock \href {https://doi.org/10.1201/b11644-11}
  {\path{doi:10.1201/b11644-11}}.

\bibitem[ZSD10]{zhu2010numerical}
H.~Zhu, H.~Shu, and M.~Ding.
\newblock {Numerical solutions of two-dimensional Burgers' equations by
  discrete Adomian decomposition method}.
\newblock {\em Computers \& Mathematics with Applications}, 60(3):840--848,
  2010.
\newblock \href {https://doi.org/10.1016/j.camwa.2010.05.031}
  {\path{doi:10.1016/j.camwa.2010.05.031}}.

\end{thebibliography}

\end{document}